\documentclass[prb,twocolumn,amsmath,amssymb,showpacs,superscriptaddress,aps,10pt,floatfix]{revtex4-1}

\usepackage{graphicx}
\usepackage{dcolumn}
\usepackage{bm}
\usepackage{multirow}
\usepackage[super]{nth}

\begin{document}


\title{Dark-field transmission electron microscopy and the Debye-Waller factor of graphene}

\author{Brian Shevitski}
\thanks{Authors contributed equally to this work.}
\affiliation{Department of Physics and Astronomy, University of California, Los Angeles, California, 90095, USA}
\affiliation{California NanoSystems Institute, University of California, Los Angeles, California, 90095, USA}

\author{Matthew Mecklenburg}
\thanks{Authors contributed equally to this work.}
\affiliation{Department of Physics and Astronomy, University of California, Los Angeles, California, 90095, USA}
\affiliation{California NanoSystems Institute, University of California, Los Angeles, California, 90095, USA}
\affiliation{Microelectronics Technology Department, The Aerospace Corporation, Los Angeles, California 90009, USA}

\author{William A. Hubbard}
\affiliation{Department of Physics and Astronomy, University of California, Los Angeles, California, 90095, USA}
\affiliation{California NanoSystems Institute, University of California, Los Angeles, California, 90095, USA}

\author{E. R. White}
\affiliation{Department of Physics and Astronomy, University of California, Los Angeles, California, 90095, USA}
\affiliation{California NanoSystems Institute, University of California, Los Angeles, California, 90095, USA}

\author{Ben Dawson}
\affiliation{Department of Physics and Nanoscience Technology Center, University of Central Florida, Orlando, FL, 32816, USA}

\author{M. S. Lodge}
\affiliation{Department of Physics and Nanoscience Technology Center, University of Central Florida, Orlando, FL, 32816, USA}

\author{Masa Ishigami}
\affiliation{Department of Physics and Nanoscience Technology Center, University of Central Florida, Orlando, FL, 32816, USA}

\author{B. C.  Regan}
\email{regan@physics.ucla.edu}
\affiliation{Department of Physics and Astronomy, University of California, Los Angeles, California, 90095, USA}
\affiliation{California NanoSystems Institute, University of California, Los Angeles, California, 90095, USA}

\date{\today}

\begin{abstract}

Graphene's structure bears on both the material's electronic properties and fundamental questions about long range order in two-dimensional crystals.  We present an analytic calculation of selected area electron diffraction from multi-layer graphene and compare it with data from samples prepared by chemical vapor deposition and mechanical exfoliation.  A single layer scatters only 0.5\% of the incident electrons, so this kinematical calculation can be considered reliable for five or fewer layers.  Dark-field transmission electron micrographs of multi-layer graphene illustrate how knowledge of the diffraction peak intensities can be applied for rapid mapping of thickness, stacking, and grain boundaries. The diffraction peak intensities also depend on the mean-square displacement of atoms from their ideal lattice locations, which is parameterized by a Debye-Waller factor.  We measure the Debye-Waller factor of a suspended monolayer of exfoliated graphene and find a result consistent with an estimate based on the Debye model. For laboratory-scale graphene samples, finite size effects are sufficient to stabilize the graphene lattice against melting, indicating that ripples in the third dimension are not necessary.

\end{abstract}

     
\maketitle

\section{Introduction}

Few-layer graphenes are quasi-two-dimensional, pure carbon materials with electronic properties that vary markedly depending on the number of layers, how they are stacked, and on defects such as grain boundaries.\cite{2009NetoReview,2011HuangMuller,2011KimZettlGrain,2012BrownMuller} One of the most effective techniques for determining these important structural characteristics is proving to be transmission electron microscopy (TEM).\cite{2007MeyerNature,2007MeyerSSC,2011KimZettlGrain,2012BrownMuller}  For samples that can be suspended or mounted on an ultra-thin, electron-transparent substrate, TEM has a unique combination of advantages. TEM can  detect three-dimensional ripples and corrugation,\cite{2007MeyerNature,2007MeyerSSC} rapidly map wide areas for grain boundaries\cite{2011HuangMuller,2011KimZettlGrain,2012BrownMuller} and thickness,\cite{2012BrownMuller} determine lattice orientation and mismatch,\cite{2011HuangMuller,2011KimZettlGrain,2012BrownMuller} and resolve stacking sequence\cite{2012BrownMuller,2012Warner} and atomic-scale defects.\cite{2011HuangMuller,2011KimZettlGrain,2012Meyer}  However, despite these capabilities the most effective use of TEM for characterizing few-layer graphene has been hindered by the lack of a quantitative, analytical model that describes how these polydisperse materials scatter electrons.

We present here a kinematical description of selected area electron diffraction from multi-layer graphene, and data which illustrate features of the calculation.  For most other materials a kinematical calculation is not practical, since the electron-crystal interaction is so strong that multiple-scattering processes cannot be neglected.\cite{2008Fultz} These non-linear effects generally require a numerical treatment. Simulations based on multislice algorithms  have been shown to be valuable tools for interpreting high-resolution TEM images of graphene, and side-by-side comparisons allow thickness and stacking determinations to be made.\cite{2010Nelson,2011Zan,2012Warner} However, because graphene consists of only a few layers of light atoms,  dynamical scattering is unimportant in the cases of most interest (layer number $N\lesssim 5$). By describing how the diffraction peak intensities scale with $N$, the relatively simple kinematical treatment offers direct prescriptions for facile grain-boundary mapping and precise layer number determination.

The complete analytic treatment includes a Debye-Waller factor, which measures long range crystalline order in terms of the mean-square displacement of atoms from their ideal lattice positions.  As was pointed out by Peierls and independently by Landau in the 1930's, the effects of thermal motion on long range crystalline order depend markedly on the dimension of the system considered.\footnote{R.E. Peierls, Helv. Phys. Acta \textbf{7}, 81 (1934) as translated in R.E. Peierls and R.H. Dalitz,  \textit{Selected scientific papers of Sir Rudolf Peierls : with commentary} (World Scientific, Singapore, 1997).}$^,$\cite{2003Landau} In one dimension long range crystalline order is precluded, while in three dimensions there is a melting temperature marking the transition between phases with and without long range crystalline order.  The study of phase transitions in the intermediate case, two dimensions, has given rise to fruitful ideas concerning topological order.\cite{1973Kosterlitz,1978Halperin}  In 1968 Mermin showed that, under quite general assumptions, crystalline order in two dimensions is excluded in the thermodynamic limit.\cite{1968Mermin}  However, he also noted that the approach to the thermodynamic limit is so slow that it might be irrelevant for practically-sized samples.  

In diffraction experiments the Debye-Waller factor describes the exponential decay of the peak amplitudes with increasing scattering angle.  Recent calculations have found that graphene's Debye-Waller factor is singular except at zero temperature,\cite{2009Tewary} which is to say that the mean-square displacements caused by thermal fluctuations become infinite in the thermodynamic limit.  This singularity is a direct manifestation of the expected lack of long range order for such a two-dimensional system. We measure the Debye-Waller factor of a suspended sample of mechanically-exfoliated monolayer graphene at room temperature, and find it to be more than twice as large as the zero temperature expectation.  Comparison with an estimate of the Debye-Waller factor based on the Debye model for the phonon band structure shows that finite size effects are sufficient to stabilize the graphene lattice.

\section{Theory}
The construction of a practical model of image formation in a modern TEM is a daunting task, since the electromagnetic lens parameters are not fixed as in an optical microscope, and the operative contrast mechanisms can vary widely depending on the microscope's imaging mode.  However, the signals generated in dark-field TEM imaging are fundamentally based on electron diffraction, and this relatively simple process can be modeled analytically.  In particular, the selected area diffraction pattern generated in a TEM is largely independent of otherwise crucial microscope parameters such as spherical aberration, electron source type, and objective lens defocus.  Dark-field images generated with a small objective aperture have signal intensities that are amenable to a quantitative, analytic description that is almost microscope independent.

\begin{figure*}\begin{center}
\includegraphics[width=\textwidth]{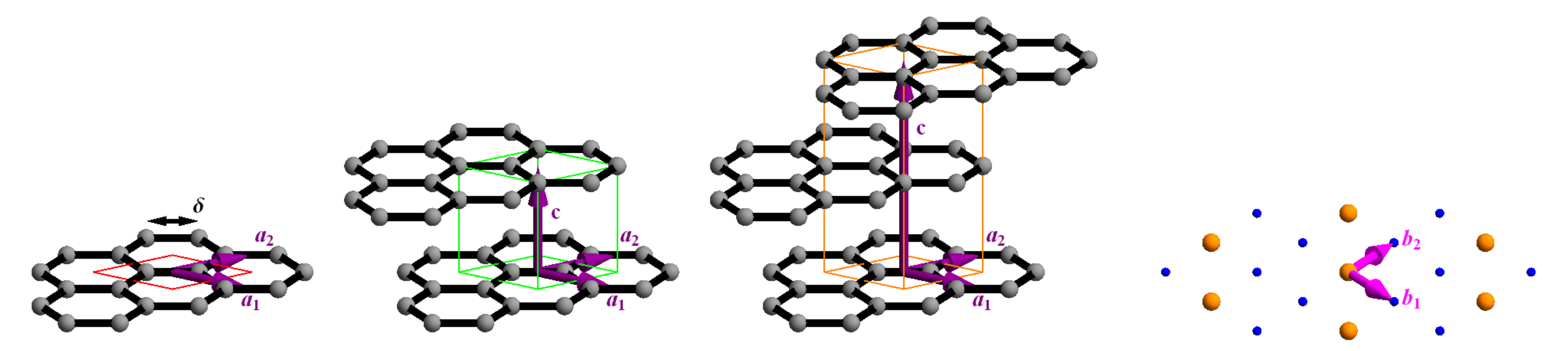}
\caption{\label{fig:crystal} (Color online) Unit cells and primitive vectors are shown for monolayer, AB-stacked bi-layer, and ABC-stacked tri-layer graphene. On the far right is the graphene reciprocal lattice, with large and small spheres representing strong and weak peaks respectively.}
\end{center}\end{figure*}

We calculate the electron diffraction pattern produced by the TEM beam using kinematical scattering theory, which has its beginnings in the integral form of the Schr\"odinger equation.  The number of electrons $dN$ scattered into solid angle $d \Omega$ is given by\cite{2005Griffiths}
\begin{equation}
\label{eq:dndo}
dN= \left(\frac{I \tau}{A e}\right)|f(\Delta \textbf{k})|^2 \, d \Omega  ,
\end{equation}
where $I$ is the beam current, $\tau$ is the exposure time, $A$ is the illuminated area, and $e$ is the electron's charge. To find the scattering amplitude $f(\Delta \textbf{k})$ we treat the crystal potential $V(r)$ as a perturbation and invoke the Born approximation. Few-layer graphene represents a nearly ideal subject for this approximation, since both carbon's small atomic number ($Z=6$) and the material's thinness act to make the scattered amplitude small. At the end of the calculation we will return to discuss the limits of the Born approximation's validity. The scattering amplitude of a crystal can be written\cite{2008Fultz}:
\begin{equation}
\label{eq:scatt}
\begin{split}
f(\Delta \textbf{k})=\sum_{\text{lattice}}e^{-i ( \Delta \textbf{k} \cdot \textbf{R}_{l})} \sum_{\text{basis}}e^{-i \left(\Delta \textbf{k} \cdot \textbf{R}_{b}\right)}\times \\ 
\left(\frac{\gamma \, m}{2 \pi \hbar^2}\right)\int e^{-i(\Delta \textbf{k} \cdot \textbf{r}^{\prime})} V(r^{\prime})d^3\textbf{r}^{\prime} \, .
\end{split}
\end{equation}
Here $\Delta \textbf{k}=\textbf{k}_f-\textbf{k}_i$ is the difference between the final and initial electron wavevectors, and $\gamma$ is the Lorentz factor (for 80 k$e$V electrons $\gamma = 1.16$).  A general position $\textbf{r}$ in the crystal has been written $\textbf{r}= \textbf{R}_{l}+\textbf{R}_{b}+\textbf{r}^{\prime}$, where $\textbf{R}_{l}$ and $\textbf{R}_{b}$ are the discrete lattice and basis vectors of the crystal, respectively.  The continuous coordinate $\textbf{r}^{\prime}$ varies over the positions nearer to one atom than to any other.

We abbreviate the scattering amplitude of Eq.~(\ref{eq:scatt}) $f(\Delta \mathbf{k}) =\mathcal{S}_l \mathcal{S}_b f_\text{atom}(\Delta k)$, which defines the lattice sum $\mathcal{S}_l$, the basis sum $\mathcal{S}_b$, and the atomic form factor $f_\text{atom}(\Delta \mathbf{k})$.  To evaluate $f_\text{atom}(\Delta \mathbf{k})$  we use the Yukawa potential to describe the screening of the nuclear Coulomb field by the atomic electrons, taking  $V(r^{\prime})=(Z e^2/4\pi \varepsilon_0 r')e^{-\mu r^{\prime}}$.  Here $\mu$ is the inverse screening length. Extending the bounds of the integral, which properly covers only a single atomic site, to include all of space introduces negligible error. In this approximation  the integral is easily evaluated, giving
\begin{equation}\label{eq:integral}
f_\text{atom}(\Delta \mathbf{k})= \left(\frac{\gamma \, m}{2 \pi \hbar^2}\right)\left(\frac{Z e^2}{4\pi \varepsilon_0}\right)\frac{4 \pi}{\mu^2+\Delta \mathbf{k}^2}.
\end{equation}
Scattering in the forward direction is clearly preferred, since the atomic form factor (\ref{eq:integral}) provides a progressive suppression as $\Delta\mathbf{k}$ becomes large.

Previous work on the TEM of few-layer graphene numerically simulates\cite{2007MeyerSSC} electron diffraction using atomic form factors that in turn are numerically estimated\cite{1968Doyle} using the relativistic Hartree-Fock atomic wavefunctions of a free carbon atom.  Expression (\ref{eq:integral}) has the advantage of being analytic with a simple physical interpretation. For a screening length $\mu^{-1} = 33$~pm Eq.~(\ref{eq:integral}) furthermore agrees with the Hartree-Fock calculation\cite{1968Doyle} for carbon to within 20\% for all $\Delta \mathbf{k}$, and to within 2\% relative to the $\Delta \mathbf{k}=0$ value $f_\text{atom}(0)=290$~pm.

The lattice sum, or shape factor, $\mathcal{S}_l$ is a geometric series that runs over the $N_c$ unit cells illuminated by the beam.  Performing the sum and squaring gives:
\begin{equation}\label{eq:sinc}
|\mathcal{S}_l(\Delta \mathbf{k})|^2= \frac{\sin^2(\frac{N_{1}}{2} \textbf{a}_{1} \cdot \Delta \textbf{k})}{\sin^2(\frac{1}{2} \textbf{a}_{1} \cdot \Delta \textbf{k})}\frac{\sin^2(\frac{N_{2}}{2} \textbf{a}_{2} \cdot \Delta \textbf{k})}{\sin^2(\frac{1}{2} \textbf{a}_{2} \cdot \Delta \textbf{k})},
\end{equation}
where  $\textbf{a}_{1}$ and $\textbf{a}_{2}$ are graphene's lattice vectors and $N_1 N_2 = N_c=A/A_\text{cell}$. The area of a unit cell $A_\text{cell}=\sqrt{3}a^2/2$ with $a \equiv |\textbf{a}_{1}|=|\textbf{a}_{2}|$. In the limit $N_c\rightarrow \infty$, the square of the shape factor is proportional to a comb of Dirac $\delta$-functions, since\footnote{G. B. Arfken, H. J. Weber, and F. E. Harris, \emph{Mathematical methods for physicists : a comprehensive guide} (Academic Press, Amsterdam, 2013) contains an error in its discussion of Fejer's method.}
\begin{equation}
\label{sincrel}
\lim_{\text{n} \to \infty}\frac{1}{\text{n}}\int_{-1/2}^{1/2}f(t)\frac{\sin^2(\text{n}\pi t)}{\sin^2(\pi t)}\text{dt}=f(0).
\end{equation}
For a 1~$\mu$m$^2$ illuminated area $N_c \simeq 2\times 10^7$, so the infinite limit is typically a good approximation here.  Writing the wavevector difference as $\Delta\mathbf{k}=\mathbf{G}+\mathbf{g} + \Delta \mathbf{k}_\perp$, where $\mathbf{G}=v_1 \textbf{b}_1 + v_2 \textbf{b}_2$ is a (discrete) reciprocal lattice vector, $\mathbf{g}=\alpha \mathbf{b}_1 +\beta \mathbf{b}_2$ is a (continuous) vector in the Brillouin zone surrounding $\mathbf{G}$, and $\Delta \mathbf{k}_\perp$ is the component of $\Delta \mathbf{k}$ perpendicular to the sheet, gives 
\begin{equation}\label{eq:delta}
|\mathcal{S}_l(\Delta \mathbf{k})|^2\simeq \text{N}_{c}\delta(\alpha)\delta(\beta).
\end{equation}
Here we have used the orthogonality property $\mathbf{a}_i \cdot \mathbf{b}_j = 2 \pi \delta_{ij}$ relating the primitive vectors of the direct and reciprocal lattices.  

Equation (\ref{eq:delta}) is non-zero only for $\Delta \mathbf{k} = \mathbf{G}+\Delta \mathbf{k}_\perp$, so we can restrict our analysis of the basis sum $\mathcal{S}_b$ to this case.  For $N$ parallel honeycomb nets with relative displacements $\mathbf{R}_j$, the sum over the basis has the form
\begin{equation}\label{eq:basissum}
\mathcal{S}_b(\Delta \mathbf{k}) = 2 \cos \left[\frac{\pi}{3} (v_1+v_2)\right]\sum_j^N e^{-i \mathbf{R}_j\cdot \Delta \mathbf{k}}.
\end{equation}
Here we have chosen to have $\mathbf{a}_1\cdot\mathbf{a}_2=a^2 \cos(\pi/3)$ and $\mathbf{b}_1\cdot\mathbf{b}_2=b^2 \cos(2\pi/3)$, as shown in Fig.~\ref{fig:crystal}.  The reverse convention can be obtained with the substitution $v_2\rightarrow -v_2$.  

The integers $v_1$ and $v_2$ index the in-plane diffraction peaks at normal incidence.  We will refer to peaks that occur at the same radius from the reciprocal space origin as $n^\text{th}$ order, where $n$ indexes the number of different peak radii occurring between the origin and the peaks under consideration.  Thus the peaks at $|\Delta \mathbf{k}|=\frac{4 \pi}{3 \delta}\equiv G_\text{min}$ at normal incidence are ``\nth{1} order'' and those at $|\Delta \mathbf{k}|=\sqrt{3}G_\text{min}$ are ``\nth{2} order''.   Here $\delta=a/\sqrt{3}\simeq 142$~pm is the carbon-carbon bond length. (See Fig.~\ref{fig:fit}(a) for a graphical representation.) 

For monolayer graphene the sum in (\ref{eq:basissum}) gives unity and the calculation of $\mathcal{S}_b(\Delta \mathbf{k})$ is complete.  For multilayer graphene the three most symmetric stackings (shown in Fig.~\ref{fig:crystal}) are designated AA (simple hexagonal), AB (Bernal), and ABC (rhombohedral), with relative in-plane displacements of 0, $(j\mod 2) \times(\mathbf{a}_1+\mathbf{a}_2)/3$, and $(j\mod 3) \times(\mathbf{a}_1+\mathbf{a}_2)/3$ respectively.  Layer $j+1$ is displaced by $j\, c \simeq 335\,j$~pm in the plane-normal direction for $j \in [0,N-1]$. For the most common case of AB stacking, the square of Eq.~(\ref{eq:basissum}) is then:
\begin{align}\label{eq:SbSquaredAB}
|\mathcal{S}_b(\Delta \mathbf{k})|^2 &= 16 \cos^2[\frac{\pi}{3} (v_1+v_2)]\times\nonumber\\
 & \cos^2 [\frac{\pi}{3} (v_1+v_2)+ \frac{c \Delta k_\perp}{2}]\frac{\sin^2( \frac{N}{2}c \Delta k_\perp)}{\sin^2( c \Delta k_\perp )},
\end{align}
for an even number of layers $N$.  The cases of an odd number of AB layers, AA stacking, or ABC stacking present no special difficulties and give similar expressions. 

Concluding the second line of Eq.~(\ref{eq:SbSquaredAB}) is a ratio analogous to the terms appearing in the shape factor (\ref{eq:sinc}), but in the case of interest here the number of layers $N$ is of order unity so the delta function approximation is not appropriate.  Equation (\ref{eq:SbSquaredAB}) describes how the diffraction peak intensities modulate as $\Delta k_\perp$ is varied, which can be accomplished by tilting the sample in the TEM.  This modulation as a function of tilt angle has been observed previously,\cite{2007MeyerNature,2007MeyerSSC,2012BrownMuller} and provides one method for determining the number of layers present in multi-layer graphene. 

The magnitude $|\Delta \mathbf{k}_\perp|$ for a given diffraction peak varies like the sine of the tilt angle, and the sine of the angle between the tilt axis and the relevant reciprocal lattice vector $\mathbf{G}$. For the case of bilayer graphene, Eq.~(\ref{eq:SbSquaredAB}) predicts that the \nth{1} order diffraction peaks will show intensity maxima at angles as small as $\arcsin(\delta/2c)\simeq 12^\circ$ and intensity minima at angles as small as $\arcsin(\delta/4c)\simeq 6^\circ$.  A \nth{2} order peak has a local maximum at normal incidence and can go through a minimum at a tilt angle of $11^\circ$ if the corresponding $\mathbf{G}$ is orthogonal to the tilt axis.  All of these features are evident in the data of Refs.~\onlinecite{2007MeyerNature,2007MeyerSSC,2012BrownMuller}.  Because the \nth{2} order peaks are at a local maximum of Eq.~(\ref{eq:SbSquaredAB}) and the \nth{1} order peaks are not, the \nth{1} order peaks are sensitive to small variations away from normal incidence.  This sensitivity to tilt angle can be exploited to reveal twinning transitions between AB stacking and its mirror image (`AC' stacking) in Bernal stacked bilayer graphene.\cite{2012BrownMuller}

Neglecting the curvature of the Ewald sphere,  $|\Delta \mathbf{k}_\perp|\simeq 0$ at normal incidence  and the expression (\ref{eq:SbSquaredAB}) simplifies considerably.  Table~\ref{table:structure} summarizes the possible values of $|\mathcal{S}_b(\mathbf{G})|^2$ for the three common stacking types.  For ABC stacking the weak peaks have zero intensity if the number of layers $N$ is divisible by three, a feature that can be exploited in `forbidden reflection imaging' for sensitive determinations of defect and layer morphology.\cite{1974Cherns,1986Alexander} Note also that for the case of AB stacking the $|\mathcal{S}_b|^2$ are identical for single and bilayer graphene in the weak diffraction orders. (Of the first few orders, the \nth{0}, \nth{2}, and \nth{5} are strong, while the \nth{1}, \nth{3}, and \nth{4} are weak.  See Table~\ref{table:structure} and Fig.~\ref{fig:fit}.) This degeneracy can have experimental consequences.  

As demonstrated previously,\cite{2011HuangMuller, 2011KimZettlGrain,2012BrownMuller} grain boundaries in polycrystalline graphene can be mapped in dark-field TEM by placing an aperture in the back focal plane to select a certain diffraction peak.  If the \nth{1} order peaks are chosen to perform the mapping as in Refs.~\onlinecite{2011HuangMuller, 2011KimZettlGrain}, then single and bilayer graphene (and two-thirds of the ABC-stacked graphenes) give indistinguishable intensities.  However, in \nth{2} order the simple $N^2$ scaling breaks these degeneracies,  allowing the facile visualization of changes in the thickness of few-layer graphene.  Furthermore, the larger prefactor in \nth{2} order means that these peaks also provide more signal in all cases except monolayer graphene.  We will see that for monolayer graphene the \nth{2} order peaks are expected to be 5\% dimmer than the \nth{1} order peaks due to the suppression caused by the atomic form and Debye-Waller factors, but, because of the decreased background at larger $\Delta k$, the signal to background ratio is still improved in \nth{2} order in all but the cleanest samples.

{\renewcommand{\arraystretch}{1.3}
\renewcommand{\tabcolsep}{0.2cm}
\begin{table}
		\begin{tabular}{| c|| c |c| c| }\hline
      Stacking & Condition & weak order & strong order \\ \hline \hline
     AA & n.a. & $N^2$ & \multirow{5}{*}{$4 N^2$}  \\ \cline{1-3}
   \multirow{2}{*}{AB} & N is odd & $(N^2+3)/4$ &   \\
      & N is even & $N^2/4$ &  \\ \cline{1-3}
    \multirow{2}{*}{ABC} & N mod 3 $\neq 0$ & 1&  \\ 
     & N mod 3 = 0 &  0&  \\ \hline
		\end{tabular}
	\caption{Summary of $|\mathcal{S}_b|^2$, the square of the relative structure factor, for various graphene stackings. The designations `weak' (\emph{e.g.} \nth{1}) and `strong' (\emph{e.g.} \nth{2}) order are shorthand for $(v_1+v_2)\mod 3 \ne 0$ and $=0$ respectively.}	\label{table:structure}
\end{table}}

To complete the calculation we integrate Eq.~(\ref{eq:dndo}) over the Brillouin zone surrounding each reciprocal lattice vector $\mathbf{G}$, which gives the number of electrons scattered into each diffraction peak.  It is convenient to do this integral in terms of the reciprocal space coordinate pair $(\alpha, \beta)$, since the shape factor (\ref{eq:delta}) is written in terms of $\delta$-functions over these variables.   Jacobian transformation of the differential solid angle element gives $d \Omega = \frac{2}{\sqrt{3}} (\lambda/a)^2 d\alpha d\beta$, where $\lambda = 2 \pi/k$ is the electron wavelength and we have taken the scattering angle to be small. (For an 80~kV accelerating potential $\lambda\simeq 4.2$~pm.) Collecting our results for the normal incidence case gives, for the number of electrons $N_\text{peak}$ scattered into a peak at $\mathbf{G}$,
\begin{equation}\label{eq:final}
N_\text{peak}=\frac{16}{27}\frac{I \tau}{e} \left( \frac{Z \gamma \lambda}{a_B }\right)^2 |\mathcal{S}_b(v_1,v_2)|^2 \frac{e^{-2W}}{\delta^4\left(\mu^2+\Delta\textbf{k}^2\right)^2}.
\end{equation}
Here we have defined the Bohr radius $a_B=\hbar/(\alpha_c m c)\simeq 53$~pm and included a Debye-Waller factor $e^{-2W}$. To an excellent approximation $\Delta \textbf{k}^2\simeq G^2=(4 \pi/3\delta)^2(v_1^2+v_2^2-v_1 v_2)$, but since
\begin{equation}\label{eq:excitation}
\Delta \textbf{k}^2 = 2 k^2 \left(1-\sqrt{1-(G/k)^2}\right),
\end{equation}
is exact (at normal incidence) we use it for fitting.

The result (\ref{eq:final}) has one great advantage that might not have been expected at the beginning of the calculation, which is that $N_\text{peak}$ only depends on the microscope settings through $I$, $\gamma$, and $\lambda$.  Together these variables do no more than modify the scale factor multiplying the entire expression.  Thus Eq.~(\ref{eq:final}) does not require special calibration measurements and can easily be applied to fit diffraction data acquired in any TEM, as we will show later.

The kinematical diffraction theory developed here is not generally considered reliable for crystals,\cite{2008Fultz} since the cumulative effect of many scattering centers (see the $N_c$ in Eq.~\ref{eq:delta}) can make the resultant wavefunction quite different from the unperturbed wavefunction.  For $N$-layer graphene, however, we can systematically evaluate the limits of the kinematical treatment. Equation~(\ref{eq:final}), when summed over all $v_1$ and $v_2$, gives the total number of electrons scattered from the TEM beam by multilayer graphene.  The first order Born approximation is valid to the extent that this number is small compared to the number of incident electrons $I \tau/e$.  The sum over diffraction peaks converges rapidly and can be performed numerically. Taking the Debye-Waller factor $e^{-2W}=1$ so as to include electrons scattered coherently into the thermal diffuse background,\cite{2008Fultz} we find that a single graphene layer scatters 0.45\% of the incident beam for an accelerating voltage of 80~kV.  This fraction includes those `scattered' into the \nth{0} order $v_1=v_2=0$ peak, which represent almost half (46\%) of the total number scattered. The total fraction scattered increases roughly like $N^2$, as shown in Table~\ref{table:structure}. For AB stacking the fraction scattered is 9\% for $N=5$ layers and 36\% for $N=10$ layers.  For $N=17$ layers the first order Born approximation has broken down entirely, since the scattered beam is implied to contain more electrons than the incident beam.  Thus we expect this kinematical theory to give excellent-to-good quantitative agreement for $N=1$--$5$ layers, and reasonable qualitative predictions for $N\lesssim 10$ layers. Quantitative analysis of thicker graphenes requires dynamical diffraction theory.

In Eq.~(\ref{eq:final}) the Debye-Waller factor $e^{-2W}$ arises as a result of disorder in the lattice, and decreases the number of electrons diffracted into a given peak. In graphene this factor touches on famous old questions about the fundamental stability of two-dimensional crystals.\cite{2003Landau,2007MeyerNature} In an infinite, two-dimensional crystal at finite temperature, thermal fluctuations create a divergence in $2W$.  For an anisotropic, layered material like graphene the Debye-Waller exponent can be written,
\begin{equation}
2W=\Delta\mathbf{k}_\text{p}^2 u_\text{p}^2 + \Delta\mathbf{k}_\perp^2 u_\perp^2,
\end{equation} 
where the $u$'s refer to the mean-square displacements of unit cells from their ideal lattice positions, and the subscripts $\text{p}$ and $\perp$ designate the in-plane and normal components respectively.  For graphene specifically the mean-square displacements have been recently calculated,\cite{2009Tewary} with the results $u_\text{p}^2=16$~pm$^2$ and $u_\perp^2=40$~pm$^2$ at $T=0$. As a particular example of the classic two-dimensional problem, graphene's Debye-Waller exponent is found to be singular at non-zero temperatures. A qualitative analysis valid for temperatures $\lesssim 1$~K indicates that finite size effects remove the singularity, but no numerical estimates more precise than the $T=0$ values for the mean-square displacements are given.\cite{2009Tewary}

To provide an analytic expression for the Debye-Waller factor that is valid for $T\ne 0$, we  calculate $u_\text{p}^2$ within the Debye approximation $E=\hbar v_s k$ for the phonon band structure. Since the Debye model is best in the infrared limit where the singularity occurs, we expect this approximation to successfully capture the essential physics. For $\Delta \mathbf{k}\simeq \mathbf{G}$, which corresponds to the normal incidence case of most interest here, we find
\begin{equation}\label{eq:ourDebyeWaller}
2W=2\frac{G^2}{k_D^2}\frac{k_B T}{M v_s^2}\left(\frac{x_D-x_s}{2}+\ln\left[\frac{1-e^{-x_D}}{1-e^{-x_s}}\right]\right).
\end{equation}
Here $k_B T$ is the thermal energy, $M$ is the mass of a carbon atom, $v_s$ is graphene's in-plane speed of sound, $k_D^2= 8\sqrt{3}\pi/9\delta^2$ is the square of the Debye wavevector, and $x_D=\hbar v_s k_D/k_B T$ is the ratio of the Debye and thermal energies.  A similar ratio $x_s=\hbar v_s k_s/k_B T$ is defined in terms of the smallest wavevector $k_s$ that can be supported by the crystal.  In the limit that $k_s\rightarrow 0$, Eq.~(\ref{eq:ourDebyeWaller}) illustrates the essential features of the singularity in $2W$ that occurs in two dimensions: $u_\text{p}^2$ diverges logarithmically for $T\ne 0$, but is finite at $T=0$. 

In a real crystal the wavevector $k_s= 2\pi/L$ is limited by the size $L$ of the crystal.  Since the wavevector cannot become arbitrarily small, the divergence in the Debye-Waller exponent is effectively regulated. At room temperature $x_D\simeq 9$ and $x_s\simeq 3.5\times 10^{-4}$ for a $L=10\,\mu$m crystal. Dropping the negligible terms in Eq.~(\ref{eq:ourDebyeWaller}) leaves
\begin{equation}\label{eq:ourDebyeWallerApprox}
2W\simeq G^2\left(\frac{\hbar}{k_D M v_s}+\frac{2 k_B T}{k_D^2 M v_s^2}\ln\left[\frac{k_B T}{\hbar v_s k_s}\right]\right).
\end{equation}
The first term in Eq.~(\ref{eq:ourDebyeWallerApprox}) is size and temperature-independent, and represents the contribution of zero-point motion to the Debye-Waller factor. It is also proportional to $\hbar$, which highlights its quantum origin. Numerically  the $T=0$ mean-squared displacement $u_\text{p}^2=\hbar/(k_D M v_s)=16$~pm$^2$ for $v_s=2.2 \times 10^4$~m/s, in agreement with the calculation of Ref.~\onlinecite{2009Tewary}.  The carbon atoms' zero-point motion in graphene is about equal to $\lambda$ for 80~kV electrons.

The second term diverges logarithmically as the crystal size $L\rightarrow \infty$, and increases like $T\ln T$ with temperature.  For experimentally realistic values $L=10\, \mu$m and $T=300$~K, the mean-square displacement $u_\text{p}^2=44$~pm$^2$, which is to say that the finite temperature correction at room temperature is twice the size of the $T=0$ value. For effective sizes $L$ ranging from 100~nm to 1~cm the mean-square displacement ranges from 28 to 69~pm$^2$.  

The size of this correction is remarkable: it is not overwhelmingly dominant, as might be expected for a formally divergent term, nor is it small compared to the zero point motion, even though room temperature is low compared to the in-plane Debye temperature $\Theta_D=\hbar v_s k_D/k_B \simeq 2600$~K. (In contrast the room temperature correction to the $T=0$ value of graphene's knock-on displacement cross section is tiny.\cite{2012Meyer})   Furthermore, the divergence with crystal size occurs extremely slowly, as suggested in Refs.~\onlinecite{1968Mermin,1979Peierls}.  While at $T=300$~K the root mean-square displacement $\sqrt{u_\text{p}^2}$ is almost 5\% of the carbon-carbon bond length $\delta$ for a $L=10\, \mu$m crystal, it is still less than 10\% of $\delta$ for an astronomical $L=10^{12}$~m. Thus, corrugations or ripples in the third dimension\cite{2007MeyerNature,2007MeyerSSC,2007Fasolino} are not necessary to explain the evident thermodynamic stability of laboratory-scale samples of graphene at room temperature. Surprisingly, in this case even a mole of carbon atoms, representing a crystal with linear dimension $L=126$~m, is not enough to approximate the thermodynamic limit.

Equations~(\ref{eq:ourDebyeWaller}--\ref{eq:ourDebyeWallerApprox}) imply a melting temperature $T_\text{m}$ for graphene that is weakly size-dependent. While the Lindemann melting criterion can be reformulated in two dimensions to circumvent the logarithmic divergence\cite{2000Zahn,2009Dietel}, such steps are not required for a finite-sized crystal. Choosing a standard value\cite{2007Chakravarty} for the Lindemann parameter $\mathcal{L}=\sqrt{u_\text{p}^2}/\delta$ of 15\%, we find $T_\text{m} \sim 3800$~K for an $L=10\,\mu$m graphene crystal.  Considering the uncertainty in $\mathcal{L}$, this reasonable estimate encourages confidence in the model underlying Eqs.~(\ref{eq:ourDebyeWaller}--\ref{eq:ourDebyeWallerApprox}).  

\section{Dark-field imaging}

\begin{figure*}\begin{center}
\includegraphics[width=\textwidth]{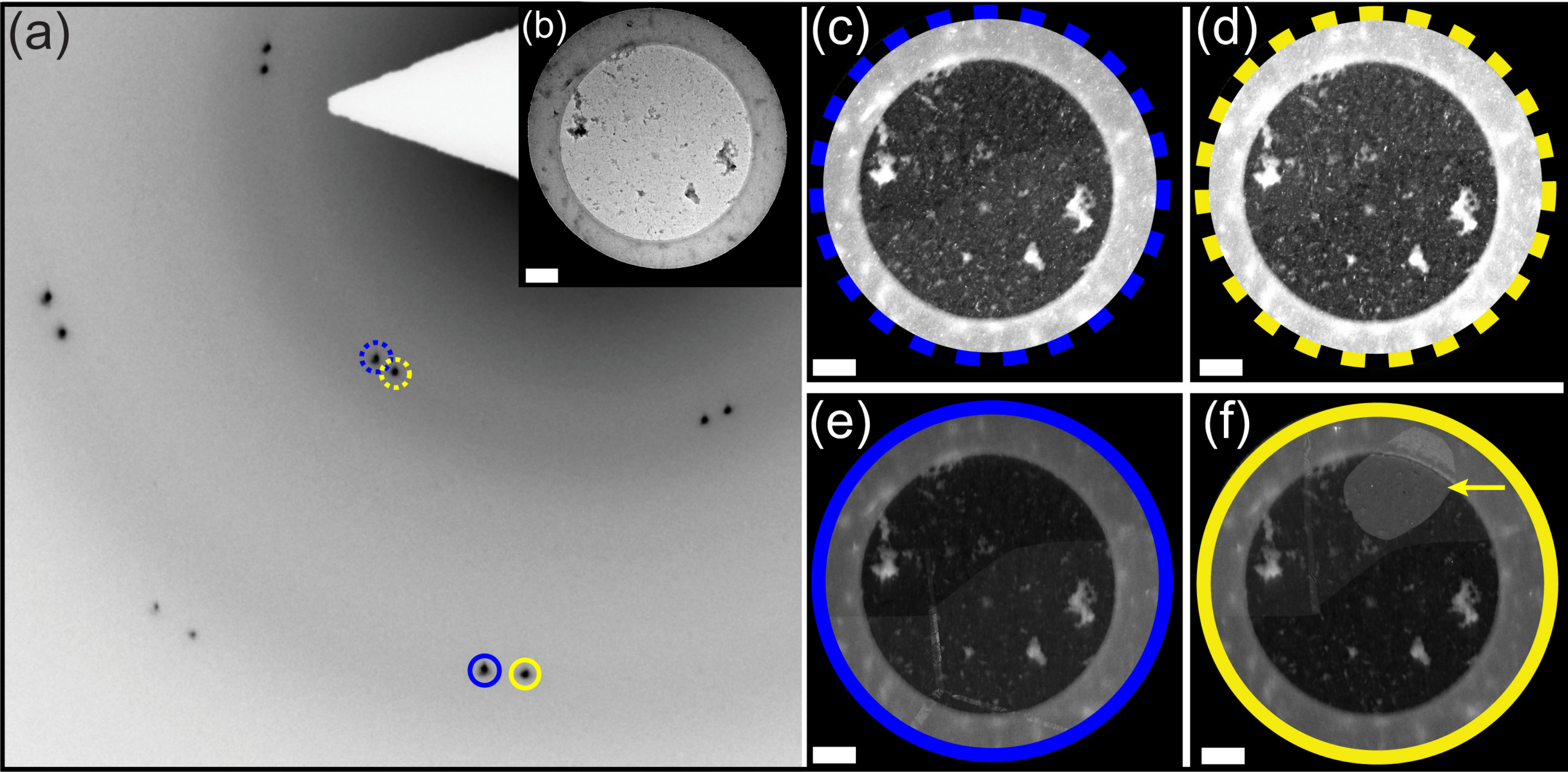}
\caption{\label{fig:cvd} (Color online) The selected area diffraction pattern (a) from CVD graphene suspended over a nominally 3~$\mu$m hole reveals the presence of two distinct grains. No grain structure is visible in the bright field image (b), where most of the contrast is generated by residual PMMA. Dark-field images (c--f) are generated using an objective aperture to separately select both the \nth{1} and \nth{2} order diffraction peaks, as indicated by dashed and solid circles on the diffraction pattern (a) respectively. All four dark-field images (c--f) were acquired with the same exposure time (5~s), and are displayed using identical contrast scales to permit fair comparison. In the images (c--d) acquired with the \nth{1} order peaks the grain boundary is barely visible.  The images (e--f) acquired with the \nth{2} order peaks show much better contrast with obvious moir\'e substructure. Furthermore, image (f) clearly reveals a lattice-aligned bilayer island (arrow) that is not visible in image (d).  The scale bars represent 500~nm.}
\end{center}\end{figure*}

The theory presented has been developed to aid our understanding of dark-field TEM images of few-layer graphene, and to help refine our procedures for acquiring such images.  In this section we present TEM images of graphene grown by chemical vapor deposition, and images of graphene prepared by the mechanical exfoliation of natural Kish graphite.  Since exfoliated graphene more closely approaches an ideal, defect-free system, it provides a basis for quantitatively evaluating the successes and failures of the theory.  CVD graphene samples represent a more polydisperse set, and here illustrate how dark-field TEM can most effectively characterize variables of paramount experimental interest, such as layer number and grain boundary structure.

To prepare CVD samples we first grow graphene on copper foil (99.8\% pure, 25~$\mu$m thick) from methane feedstock, following a procedure described previously.\cite{2009LiRuoff}  The growth occurs at $1045^\circ$C in a quartz-tube furnace with a base pressure of 10~mTorr.  To transfer the graphene onto TEM grids, we spin-coat a 300 nm thick polymethyl methacrylate (PMMA) support layer onto the graphene and etch away the copper foil with a 1.6~M solution of FeCl$_2$. After rinsing in de-ionized water the graphene/PMMA stack is then scooped onto holey carbon TEM grids and baked at 70$^\circ$C for 1 hour to remove the water. The PMMA is removed in an acetone bath, and the grid is finally rinsed with isopropanol.

We use the method presented previously \cite{2011HuangMuller, 2011KimZettlGrain, 2012BrownMuller, 2012Wassei} to determine the grain structure of a small suspended region. Briefly, we locate the region of interest using bright field TEM and acquire the selected area electron diffraction (SAED) pattern. Placing a 10~$\mu$m objective aperture in the back focal plane, we select a certain diffraction peak (or peaks) and return the microscope to imaging mode. The aperture acts as an electron filter in reciprocal space, giving a resultant real-space image with signal only from the crystallographic orientations corresponding to the diffraction peaks selected. 

Figure~\ref{fig:cvd} shows representative data acquired in the process of generating dark-field TEM images of CVD graphene. The diffraction pattern (a) shows two copies of the expected hexagonal structure with a relative rotation of $4^\circ$, indicating that the field of view  contains at least one grain boundary.  The bright field image (b) shows no graphene-specific features, but the dark-field images (c--f) map the two grains present and identify each with its specific crystallographic orientation.  We have found that the \nth{2} order peaks are generally preferred for rapid sample characterization.  When the \nth{1} order peaks are selected to generate the image the contrast between grains is low, to the point that the gross grain structure can barely be discerned without image processing.  While grain boundaries such as those  in (c--d) are perceptible, the lack of native contrast makes colorizing or boundary `guides for the eye' necessary for many display purposes.\cite{2011HuangMuller, 2011KimZettlGrain}

In comparison the grain boundaries seen in images (e--f) of Fig.~\ref{fig:cvd}, which are acquired using \nth{2} order peaks, are unmistakable.  The analysis of the preceding section  explains why better contrast is expected from these peaks.  While the atomic form factor and the Debye-Waller factor in \nth{2} order are 0.26 and 0.93 of the \nth{1} order factors respectively, the $|\mathcal{S}_b|^2$ factor is four times larger, for a net change of only $-5\%$.  However, the diffuse background in these images decreases by a factor of $\sim 2.8$ as the distance from the reciprocal space origin increases by a factor of $\sqrt{3}$, leading to the improved signal-to-background in \nth{2} order.  The $|\mathcal{S}_b|^2$ values given in Table~\ref{table:structure} also establish that the island visible in Fig.~\ref{fig:cvd}(f) is AB-stacked bilayer graphene, for (assuming $N<4$) only AB-stacked bilayer has the degeneracy required to make this island invisible in \nth{1} order (Fig.~\ref{fig:cvd}(d)).   For samples containing both single and AB-stacked bilayer graphene, comparing the \nth{1} and \nth{2} order dark-field images allows faster and more conclusive real-space layer mapping than, for instance,  acquiring a diffraction pattern tilt series for a sequence of select areas.

The \nth{2} order peaks also clearly reveal moir\'e substructure that is less evident in the \nth{1} order images (c--d) of Fig.~\ref{fig:cvd}.  The upper grain (f) is split by one vertical moir\'e stripe about 50~nm wide, and the lower grain (e) shows a 3-way intersection of similar stripes, the vertical leg of which extends into the moir\'e stripe of the upper grain.  Previously similar features have been identified in Bernal-stacked bilayer graphene, and associated with twin boundaries between AB- and AC-stacking.\cite{2012BrownMuller}  Following this suggestion we attribute the stripes seen here to narrow regions of bilayer overlap, with the 3-way intersection representing the boundary between \mbox{AB-,} \mbox{AC-,} and BC-stacked regions.

\begin{figure*}[t!]
\begin{center}
\includegraphics[width=0.9\textwidth]{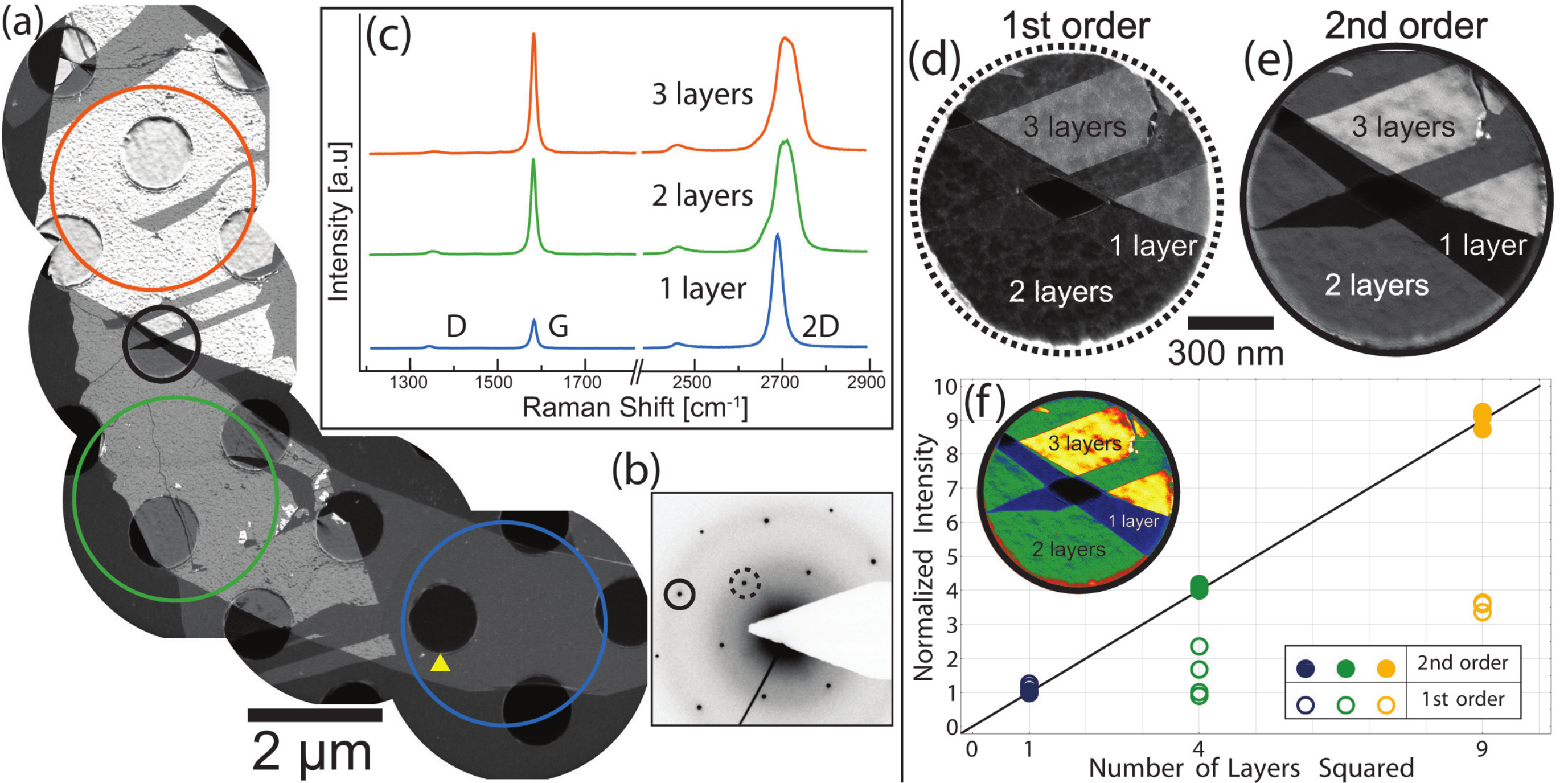}
\caption{\label{fig:exfol} (Color online) A composite dark-field image (a) of a single graphene crystal on a holey carbon grid is formed using electrons from the \nth{2} order peak, which is indicated by the solid circle on the diffraction pattern (b).  Graphene over a hole, such as the one indicated by the yellow triangle in (a), is fully suspended. Raman spectra (c), which were acquired from the regions designated by the colored circles in (a) before transfer to the grid, confirm the thickness of each region. Higher-magnification dark-field images (d--e) of the region within the black circle in (a) are acquired using the \nth{1} and \nth{2} order peaks respectively.  The same area is shown again as a false color legend in the plot (f), which depicts the average normalized intensity of each disconnected region of (d) and (e) as a function of the square of the number of layers.}
\end{center}\end{figure*}

CVD graphene shows a wide array of morphologies; a given sample can show various numbers of layers, different stackings, and non-trivial grain boundary structures.  To quantitatively test the analytic diffraction theory developed here it is preferable to work with the more homogeneous and defect-free samples that can be obtained by mechanically exfoliating natural Kish graphite.  We characterize flakes of mechanically exfoliated graphene using Raman spectroscopy, and then transfer these flakes to TEM grids for dark-field imaging.  The graphene is first deposited on 300~nm SiO$_2$ supported by a silicon substrate. The flake thickness is determined in a Raman microscope (Renishaw InVia) using a 514~nm argon excitation laser with a $\sim 4$~$\mu$m spot. After characterization, isopropanol is dripped onto the chip until a droplet covers its entire surface. A 200 mesh copper TEM grid covered by a holey carbon mesh (Protochips C-flat $1\,\mu$m holes, 2~$\mu$m pitch) is placed inside the droplet with the holey carbon facing the target flake. Once the grid is aligned with the target flake the isopropanol is allowed to evaporate, which pulls the holey carbon into contact with the graphene. Isopropanol is then added to separate the grid from the chip, and the grid is allowed to dry in air.  We find that this procedure reliably provides Raman-characterized graphene flakes suspended on a TEM grid.

Figure \ref{fig:exfol}(a) shows a composite image of a graphene flake prepared with this procedure. The flake consists of regions of varying thicknesses, as is clearly evident from intensity variations across the composite image.  This variation proportional to $N^2$ is expected, since the dark-field image is constructed using the \nth{2} order peak indicated on the diffraction pattern (b).  Raman spectra (c) collected from regions marked approximately by the colored circles in the composite (a) indicate \mbox{mono-,} \mbox{bi-,} and tri-layer graphene; the monolayer graphene has an approximately 4:1 intensity ratio between the $2D$ peak at $\sim 2700$~cm$^{-1}$ and the $G$ peak at $\sim 1580$~cm$^{-1}$, while between bi- and tri-layer graphene the ratio goes from slightly more to slightly less than one, with characteristic shoulders on the $2D$ peak.\cite{2006Ferrari,2009Malard}  A small defect, or `$D$', peak at 1350~cm$^{-1}$ is observed because the laser spot is wider than the flake.  Allowing for the signal from the graphene edges, the small size of this peak indicates that the sample is high-quality graphene relatively free of defects.

One hole in the holey carbon grid, encircled with a black ring in Fig.~\ref{fig:exfol}(a), contains suspended \mbox{mono-,} \mbox{bi-,} and tri-layer graphene together, along with an actual puncture that has no spanning membrane at all.  Higher magnification images (d--e) of this region, acquired in dark-field from the \nth{1} and \nth{2} order spots (designated with dashed and solid rings on the diffraction pattern (b)), illustrate the relative merits of these two orders for rapid sample characterization.  In the \nth{1} order (d) image the contrast change between the mono- and bi-layer graphene is negligible, as expected given the degeneracy in the basis sum $|\mathcal{S}_b|^2$. In the \nth{2} order image (e) there is a clear contrast progression as the number of layers $N$ ranges from 0 to 3.  The plot (f) shows normalized, average signal intensities collected from regions of differing thicknesses (indicated with false color in the inset). A small background, probably due to residue from the transfer process, has been subtracted from each intensity before normalization. The normalized intensities collected in \nth{2} order quantitatively illustrate the $N^2$ scaling predicted for the strong diffraction orders (see Table~\ref{table:structure}).    The dark-field intensities measured using the \nth{1} order peaks show the expected degeneracy between mono- and bi-layer graphene, along with a signal that is approximately three times larger for tri-layer. Again, these behaviors are in accordance with the theoretical results summarized in Table~\ref{table:structure}.

\section{Measurement of the Debye-Waller factor}

\begin{figure}\begin{center}
\includegraphics[width=\columnwidth]{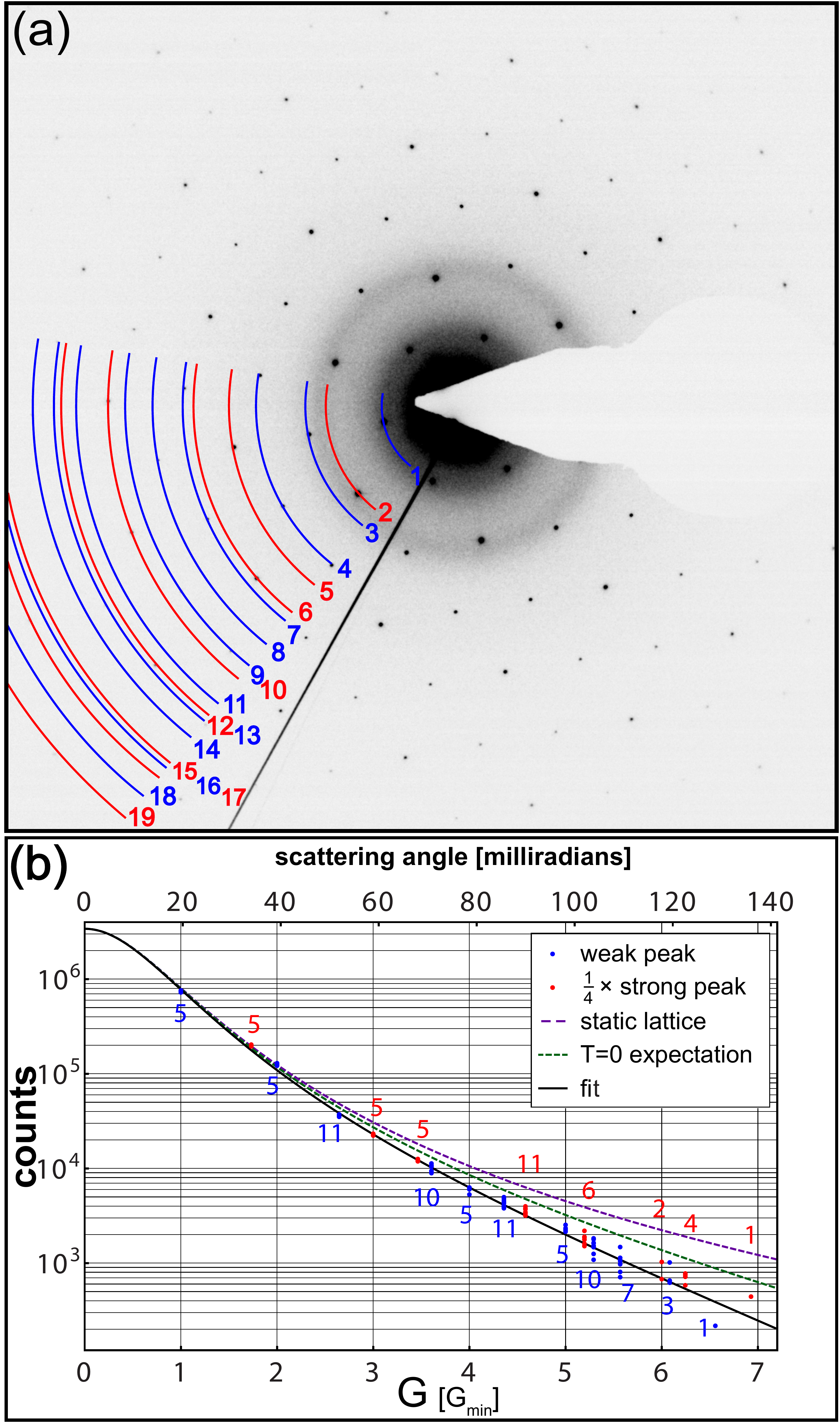}
\caption{\label{fig:fit} (Color online) The diffraction pattern (a) was acquired from the monolayer graphene suspended over the hole indicated by the yellow triangle in Fig.~\ref{fig:exfol}(a).  The integrated intensity of each diffraction peak is fit to Eq.~(\ref{eq:final}), where the strong peaks have been normalized by dividing by a factor of 4 (see Table~\ref{table:structure}).  The peak intensities diminish more quickly with increasing $G$ than can be explained by the atomic form factor (static lattice) and zero-point motion ($T=0$ expectation) alone. The number of measured peaks of each order is indicated at each radius.}
\end{center}\end{figure}

Suspended over the 1~$\mu$m hole marked with a yellow triangle in Fig.~\ref{fig:exfol}(a) is a single layer of mechanically exfoliated graphene.  We perform electron diffraction on this region to quantitatively test the analytic model summarized by Eq.~(\ref{eq:final}).  A representative diffraction pattern, shown in Fig.~\ref{fig:fit}, is acquired with a 0.4~s exposure at a camera length of 130~mm using a 2048~pixel $\times$ 2048 pixel Gatan UltraScan camera with 16 bit digitization. Dozens of diffraction patterns acquired under similar imaging conditions during separate microscope sessions give consistent results, in keeping with the expectation that the 80~kV electron beam induces negligible damage in the sample.\cite{2012Meyer} The beam blocker prevents measurement of some peaks, but roughly 5/6$^\text{ths}$ of the total number are visible.  The ratio of the average intensity of the visible \nth{2} order peaks to that of the \nth{1} order peaks is 1.05.  Since the expected ratios are 0.95, 3.8, 2.9, 3.8, and 3.4 for 1--5 AB-stacked layers respectively, this measurement further establishes the material as monolayer graphene.

We now perform a fit to Eq.~(\ref{eq:final}) on 112 peaks extracted from the diffraction pattern in Fig.~\ref{fig:fit}(a). Each peak is fit individually to a 2D Gaussian with nine free parameters: an amplitude, a 2D linear offset background, two center coordinates, two widths, and an angle describing the relative rotation between the Gaussian axes and the diffraction image axes. All points in the fit region are weighted equally. A typical width is $\sim 2$~pixels, the separation between peaks is 189~pixels, and a fit region is 60~pixels $\times$ 60~pixels, so the offset gives a good measure of the non-Bragg scattering background in the neighborhood of the specific peak. Near the reciprocal space origin the measured peaks have an excess in the wings relative to the Gaussian fits, so a peak intensity is calculated by summing the counts, minus the offset background, in a 20~pixel $\times$ 20~pixel region about the Gaussian center.  This procedure gives peak intensities that are insensitive to the details of the fitting function, except through its determination of the background.  These intensities are plotted in Fig.~\ref{fig:fit}(b) as a function of distance $G=|\textbf{G}|$ from the reciprocal space origin, or, equivalently, the scattering angle.  The strong peaks have been normalized by dividing the integrated intensity by 4. 

We fit peaks as far out as \nth{19} order, which is at a distance $4\sqrt{3} G_\text{min}$ from the origin.  The standard deviation of the intensities at each diffraction order is taken as an estimate of the errors used to provide weights for the fit.  The fit has three free parameters: a scale factor proportional to the (unmeasured) beam current $I$, the inverse screening length $\mu$, and the mean-square displacement $u_\text{p}^2$, since the other parameters in Eq.~(\ref{eq:final}) are well known. The results are insensitive to strain, since uniform strain only shifts the measured peak positions (which are not used) and non-uniform strain only broadens the peaks\cite{2008Fultz} (which does not change their integrated intensities). Note also that most of the effects of the curvature of the Ewald sphere, \emph{e.g.} the excitation error, are included through use of Eq.~(\ref{eq:excitation}). As is evident from Eq.~(\ref{eq:scatt}), for monolayer graphene the perpendicular component of $\Delta \textbf{k}$ only enters the scattering amplitude through the atomic form factor (\ref{eq:integral}). 

The final result is shown in Fig.~\ref{fig:fit}(b), along with static lattice and zero temperature predictions obtained from Eq.~(\ref{eq:ourDebyeWaller}).  For the screening length we find $\mu^{-1}=34 \pm 2$~pm, in agreement with our estimate of 33~pm based on the results of the relativistic Hartree-Fock calculation in Ref.~\onlinecite{1968Doyle}.  For the mean-square displacement we find $u_\text{p}^2=40\pm 10$~pm$^2$, where the quoted uncertainty reflects a systematic influence of the camera length on the measured $u_\text{p}^2$ that is not presently understood.  This value, while significantly larger than the $T=0$ expectation of 16~pm$^2$, is consistent with the model described by Eqs.~(\ref{eq:ourDebyeWaller}--\ref{eq:ourDebyeWallerApprox}) with a characteristic crystal size $L\sim 3.5$~$\mu$m at $T=300$~K, in accord with the sample dimensions evident in Fig.~\ref{fig:exfol}(a). 

Taking the results of the fit at face value,  9\% of the electrons that would be scattered into diffraction peaks by an ideal, motionless monolayer of graphene are instead scattered into the thermal diffuse background at room temperature, as compared to 5\% due to the $T=0$ zero-point motion alone.  While the agreement between the observed size of the measured crystallite and the value implied by the fit is good, the roles of vibrations, the support structure, substrate interactions, defects and other temperature-independent disorder in the sample, and the actual phonon band structure are not yet understood.  Future studies that look at the Debye-Waller factor as a function of temperature, crystallite size, and defect density (\emph{e.g.} CVD versus mechanically exfoliated graphene) will provide further insight into the nature of long range order in this model two-dimensional system.

\section{Conclusion}

We have calculated electron diffraction from multilayer graphene using first-principles kinematical scattering theory, and presented experimental data from both CVD and mechanically exfoliated graphene that illustrate features of the calculation.  For a number of layers $N\leq 5$ the fraction of incident 80~k$e$V electrons scattered is less than 10\%, indicating that the kinematical theory is reliable in this few layer limit but that dynamical effects will become increasingly important for thicker graphenes. Expressions describing the structure factor as a function of $N$ suggest that the \nth{2} order peaks are the most useful for rapid dark-field characterization of few-layer samples.  For the purposes of mapping grain boundaries, the \nth{2} order peaks generally give better contrast.  For samples containing regions of various thicknesses, the \nth{2} order peaks give reliable thickness contrast $\propto N^2$ and in some cases allow immediate determination of the absolute number of layers.  Finally, the complete model includes an analytic calculation of the Debye-Waller factor that is valid at $T\ne0$ for samples of finite size.  This part of the calculation exemplifies a famous result for 2D systems that has been taken to imply that a 2D crystal is unstable to thermal fluctuations. Surprisingly we find that, for graphene crystals of experimentally relevant sizes at room temperature, the thermal corrections to the $T=0$ result are several times larger than the $T=0$ result itself, even though room temperature is only 10\% of graphene's Debye temperature.  We measure the Debye-Waller factor of a monolayer of mechanically exfoliated graphene at room temperature, and find it to be more than twice the size of the $T=0$ expectation.  This measurement confirms the importance of the finite-size, finite-temperature corrections.  We anticipate that future measurements of the Debye-Waller factor in graphene as a function of defect density, crystal size, and temperature will clarify the effect of very long-wavelength phonons on graphene's strength and stability. The Debye-Waller analysis presented here (and extending the diffraction calculation to give electron intensities in an image plane) might also shed light on the Stobbs factor contrast discrepancy between HRTEM experiment and simulation.\cite{2004Howie,2009Thust,2011Forbes,2012Lee}

\section*{Acknowledgments}
This project is supported by NSF CAREER Grant 0748880, the NSF CAMP and REU programs, and NIH Award R25GM055052.   M.I. is supported by NSF Grant 0955625.  B.D. and M.L.S. are supported by the Intelligence Community Postdoctoral program. All TEM work was performed at the Electron Imaging Center for NanoMachines at the UCLA CNSI, supported by NIH Award 1S10RR23057. The work at The Aerospace Corporation was supported under its Independent Research and Development program.
\bibliography{blackgraphene}

\end{document}